\documentclass[aps,amsfonts,nofootinbib,superscriptaddress]{revtex4}
\usepackage{amsmath}

\newcommand{\beq}{\begin{equation}}
\newcommand{\eeq}{\end{equation}}
\newcommand{\bea}{\begin{eqnarray}}
\newcommand{\eea}{\end{eqnarray}}
\newcommand{\cQ}{{\mathcal{Q}}}
\newcommand{\cW}{{\mathcal{W}}}
\newcommand{\dk}{{\int\frac{\d^3 \vc{k}}{(2\pi)^{3/2}}}}
\newcommand{\te}{{\tilde{\epsilon}}}
\newcommand{\tet}{{\tilde{\eta}}}
\newcommand{\tetp}{{\tet^\parallel}}

\newcommand{\vc}[1]{{\textbf{\em #1}}}

\newcommand{\lh}{\left(}
\newcommand{\rh}{\right)}
\newcommand{\der}{\partial}
\renewcommand{\d}{\mathrm{d}}
\newcommand{\non}{\nonumber}
\DeclareMathSymbol{\mg}{\mathrel}{symbols}{"1D}
\newcommand{\ml}{\ll}

\newcommand{\ga}{\alpha}

\newcommand{\gd}{\delta}

\newcommand{\gz}{\zeta}

\newcommand{\gth}{\theta}

\newcommand{\gk}{\kappa}

\newcommand{\gm}{\mu}
\newcommand{\gn}{\nu}

\newcommand{\gr}{\rho}

\newcommand{\gt}{\tau}

\newcommand{\gf}{\phi}

\newcommand{\gG}{\Gamma}

\newcommand{\gL}{\Lambda}
\newcommand{\gX}{\Xi}
\newcommand{\gP}{\varPi}

\newcommand{\cC}{{\mathcal C}}
\newcommand{\cD}{{\mathcal D}}

\newcommand{\cG}{{\mathcal G}}

\newcommand{\cJ}{{\mathcal J}}

\newcommand{\cO}{{\mathcal O}}

\newcommand{\cS}{{\mathcal S}}

\newcommand{\ti}{{\tilde i}}
\newcommand{\tj}{{\tilde j}}
\newcommand{\tk}{{\tilde k}}

\newcommand{\ttt}{{\tilde t}}

\newcommand{\tx}{{\tilde x}}

\newcommand{\tD}{{\tilde D}}

\newcommand{\tge}{{\tilde\epsilon}}

\newcommand{\tget}{{\tilde\eta}}

\newcommand{\tgx}{{\tilde\xi}}

\begin{document}

\title{Non-linear perturbations in multiple-field inflation}

\author{G.I.~Rigopoulos}
\affiliation{Institute for Theoretical Physics, Utrecht University,\\
Postbus 80.195, 3508 TD Utrecht, The Netherlands}

\author{E.P.S.~Shellard}
\author{B.J.W.~van Tent}

\affiliation{Department of Applied Mathematics and Theoretical Physics, 
Centre for Mathematical Sciences,\\ 
University of Cambridge,
Wilberforce Road, Cambridge CB3 0WA, United Kingdom}

\begin{abstract}
\noindent We develop a non-linear framework for describing
long-wavelength perturbations in multiple-field inflation. The
basic variables describing inhomogeneities are defined in a
non-perturbative manner, are invariant under changes of time
slicing on large scales and include both matter and metric
perturbations. They are combinations of spatial gradients
generalising the gauge-invariant variables of linear theory.
Dynamical equations are derived and supplemented with stochastic
source terms which provide the long-wavelength initial conditions
determined from short-wavelength modes. Solutions can be readily
obtained via numerical simulations or analytic perturbative
expansions. The latter are much simpler than the usual
second-order perturbation theory. Applications are given in
a companion paper.
\end{abstract}

\maketitle

\section{Introduction}
It is a well-established fact that the universe on large scales
exhibits a high degree of uniformity. During most cosmological
eras and for a large span of length scales, it can be well
approximated by a Friedmann-Robertson-Walker (FRW) spacetime with
inhomogeneities described as small linear perturbations around the
highly symmetric background. This picture has proved particularly
relevant for the early universe, as the smallness of the cosmic
microwave background (CMB) temperature anisotropies indicates. An
extrapolation of this observational fact suggests that the use of
linear theory would also be justified during inflation when the
perturbations leading to the CMB anisotropies are thought to have
been created. So far, almost all studies of the generation and
evolution of perturbations in inflation invoke the use of linear
perturbation theory \cite{mfb}. In principle it offers a
tremendous simplification of the task of studying the true
inhomogeneous spacetime.

However, even when attention is focused on the inflationary era,
linear theory cannot be the whole picture. Since gravity is
inherently non-linear and the potential of the inflationary model
is likely to be interacting, some small non-linearity will be
endemic to the perturbations. Given the accuracy of forthcoming
CMB observations, it is worth investigating whether this
non-linearity can be observationally relevant. The characteristic
signature of non-linear effects will be deviations of the
primordial fluctuations from Gaussian statistics. In order for
this non-Gaussianity to be calculated one needs to go to second
order in perturbation theory or develop a fully non-linear
approach.

The issue of calculating non-linearity and the consequent
non-Gaussianity in the primordial universe has been attracting
increasing attention recently, although some early attempts to
calculate it can also be found in the literature
\cite{early_attempts}. A tree-level calculation with a cubic action
for the perturbations has been performed in \cite{mald} for the
case of slow-roll single-field inflation (with a similar slow-roll
calculation for more general single-field Lagrangeans given in
\cite{lidsey}). At the level of the equations of motion, various authors have
pursued perturbation theory to second order \cite{2nd_order}, with
\cite{ng_review} providing a review of these techniques. Although
interesting results can be obtained at second order, full
exploration of the system of equations suffers from great
computational complexity. On the one hand, the perturbation
equations tend to be rather cumbersome to derive. On the other
hand, gauge-invariant variables, which have proved very useful for
computations and the interpretation of results in linear theory,
are not as simple as their first-order counterparts when
second-order perturbations are considered. 

In this paper we take a different viewpoint on the study of
non-linear perturbations during inflation, systematically
formulating and extending ideas presented in \cite{gp2}, which
incorporate the general multiple-field analysis of \cite{vantent}.
We use combinations of spatial gradients to construct variables
describing the deviation from a spatially uniform spacetime and
derive equations for these variables on super-horizon scales
(section~\ref{lwsec}). These variables are defined
non-perturbatively and are invariant under changes of the time
coordinate on such scales (appendix~\ref{appendix}). They were
first used in \cite{gp}, where we derived a non-linear
generalisation of the familiar adiabatic conservation law of
linear theory. More recently, the authors of
\cite{langlois-vernizzi} used similar combinations of gradients in
the context of the covariant formalism \cite{ellis} to also derive
this conservation law. They showed the relation of these simple,
yet fully non-linear, variables to others defined in second-order
perturbation theory. An equivalent conservation law was derived in
\cite{lms} without using such gradient variables.
[Another, more recent approach in the study of non-linear perturbations can be 
found in \cite{deltaN}, where the formalism of \cite{deltaNorig} is extended 
and used.]

In order to include the continuous influx of sub-horizon
perturbations to the long-wavelength system, stochastic noise
terms are added to the long-wavelength equations. For the
definition of these noise terms a specific choice of time turns
out to be particularly convenient. Thus, we arrive at a set of
fully non-linear stochastic equations which include both matter
and metric perturbations (section~\ref{stochsec}). For actual
multiple-field calculations it is more convenient to use the
explicit field basis of \cite{vantent}. It is introduced and the
relevant equations are rewritten in terms of this basis
(section~\ref{basissec}). In section~\ref{analnumsec} we show
how the derived equations can be used to extract information about
non-linearity in inflation. A perturbative analytic approach can
be applied giving results to second order. At first order, this
perturbative expansion is equivalent to the well-known linear
gauge-invariant perturbation theory. At second order, however, it
is much simpler than the corresponding second-order approaches
pursued to date. Since the evolution equations are fully
non-linear on long wavelengths, numerical simulations can be
performed without the need for analytic approximations. In this
paper we concentrate on presenting the methods, and we stress that
no slow-roll approximation is made here. Detailed applications are
provided in separate publications \cite{sf,mf,mf2}.

\section{Long-Wavelength Approach}
\label{lwsec}

We start by considering the metric
	\beq\label{adm}
	\d s^2=-N^2(t,\vc{x})\,\d
	t^2+\mathrm{e}^{2\alpha(t,\vc{x})}h_{ij}(t,\vc{x})\,\d
	x^i\d x^j\,,
	\eeq
where we have fixed part of the gauge by setting
the shift, i.e.\ the $g_{0i}$ component, to zero. For convenience
we will keep $g_{0i}=0$ throughout the whole paper. The spatial
part of the metric has been decomposed into a spatial metric
tensor $h_{ij}$ with unit determinant and a determinant part which
plays the role of a locally defined scale factor $a(t,\vc{x})
\equiv \exp[\alpha(t,\vc{x})]$ \cite{bardeen}. The remaining
gauge freedom is encoded in the lapse function $N(t,\vc{x})$; a
choice of $N$ corresponds to a choice of time slicing for the
spacetime. The vector normal to these time slices is given by
$n_0=-N$, $n_i=0$ and their embedding in the four-dimensional
spacetime is characterised by the extrinsic curvature tensor
$K_{ij}$. We will use the tensor
	\beq
	H_{ij}=\frac{1}{3}
	\nabla_{(i}n_{j)} =\frac{1}{6N}\,\partial_t(\mathrm{e}^{2\alpha}h_{ij})\,,
	\eeq
which is
$-(1/3)K_{ij}$. The quantity $H_{ij}$ can be decomposed into a
trace and a traceless part, respectively,
	\beq\label{hubble}
	H=\frac{1}{N} \,
	\partial_t\ga\,,\qquad\qquad \bar{H}_{ij}=-\frac{1}{6N}\,
	\mathrm{e}^{2\alpha}\partial_t h_{ij}.
	\eeq

We now focus on the first approximation we will be employing: the
long wavelength approximation \cite{sb, long wavelength}. (The
second is the use of stochastic noise terms to describe the effect
of quantum fluctuations and will be discussed in the next
section.) We will be interested in length scales larger than the
comoving Hubble radius $1/(aH)$, which are called super-horizon,
because on those scales one expects non-linearities to build up.
(On sub-horizon scales the sources of non-linearity, gravity and
interacting potentials, do not play a role and linear theory is
expected to hold to high accuracy.) Consider variations over a
characteristic comoving length scale $L$. For any quantity
$F(t,\vc{x})$ constructed out of metric and matter variables
typically we will have $\der_i F = \cO(F/L)$ and $\der_t F/N =
\cO(HF)$. From this we see\footnote{As will be explained later, we
will choose a gauge in which $aH$ does not depend on $\vc{x}$. In
this case the statement $L \mg 1/(aH)$ makes sense globally and
not just locally.} that for scales $L \mg 1/(aH)$
we can expect $\left|\frac{1}{a}\der_i F \right| \ml \left| \frac{1}{N}{\der_t
F}\right|$. Hence the long-wavelength approximation means that we
can ignore second-order spatial derivatives when compared to time
derivatives.

From the long-wavelength evolution equation for $\bar{H}_{ij}$ we
find \cite{bardeen, sb} 
	\beq 
	\der_t \bar{H}^i{}_j \simeq -3NH\bar{H}^i{}_j 
	\qquad\Rightarrow\qquad
	\bar{H}^i{}_j=C^i{}_j(\vc{x})\frac{1}{a^3}\,. 
	\eeq 
Here the approximate equality implies second-order spatial gradients are
ignored. Hence, $\bar{H}^i{}_j$ is a mode that decays
exponentially fast in an inflationary universe. From now on we
will set it to zero, thereby demanding that $h_{ij}$ depends on
$\vc{x}$ only, as can be seen from equation (\ref{hubble}). In
other words, $h_{ij}$ (which contains the tensor modes) does not
participate in the long-wavelength dynamics (gravity waves freeze
on such scales). For the sake of simplicity, henceforth we will
ignore the tensor perturbations generated in inflation and
consider the spacetime to be close to flat Robertson-Walker, thus
setting $h_{ij}(\vc{x}) = \delta_{ij}$.

For the matter sector, we consider a very general inflationary era
driven by $m$ scalar fields $\gf^A$ with the energy-momentum-tensor
	\beq\label{matter}
	T_{\mu\nu}=G_{AB}\partial_{\mu}\phi^A\partial_{\nu}\phi^B
	-g_{\mu\nu}\left(\frac{1}{2}G_{AB}
	\partial^{\lambda}\phi^A\partial_{\lambda}\phi^B+V\right),
	\eeq
with $A$, $B$, etc.\ running from 1 to $m$ and $V$ the potential.
The derivative of the fields with respect to proper time is denoted by
	\beq
	\gP^A\equiv \frac{\der_t\phi^A}{N}\,.
	\eeq
Note that in (\ref{matter})
we have taken a general field metric $G_{AB}$ and all the
equations we give below will be valid for such an arbitrary
metric. We thus view the dynamics as taking place on a general
$m$-dimensional field manifold parametrised by a set of $m$
coordinate functions $\phi^A$. Hence it makes sense to define
covariant derivatives when considering spatial or temporal
dependence. For a spacetime-dependent quantity $L^A(t,\vc{x})$
which transforms as a vector in field space, we define the
covariant derivatives
	\beq\label{cov_1}
	\cD_t{L^A}=
	\partial_tL^A +\Gamma^A_{BC}\,\partial_t\phi^BL^C\,,\qquad\qquad
	\cD_i{L^A}=
	\partial_iL^A +\Gamma^A_{BC}\,\partial_i\phi^BL^C\,,
	\eeq
with
$\Gamma^A_{BC}$ the symmetric connection formed from $G_{AB}$. The
quantities $\partial_i\phi^B$ and $N\gP^B$ transform as vectors in
field space but $\phi^B$ does not. Then the long-wavelength
equations of motion for this system are \cite{sb}
	\bea\label{H_dynamics}
	\frac{dH}{dt}&=&-\frac{\gk^2}{2}N\gP_B\gP^B\,, \\
	\label{momentum_dynamics}
	{\cD_t}\gP^A&=&-3NH\gP^A-NG^{AB}V_B\,,\\
	\label{friedmann}
	H^2&=&\frac{\gk^2}{3}\left(\frac{1}{2}\gP_B\gP^B+V\right)\,,\\
	\label{0i} 
	\partial_iH&=&-\frac{\gk^2}{2}\gP_B\partial_i\phi^B\,,
	\eea
where $V_B\equiv\partial_BV \equiv \der V / \der \gf^B$ and
$\gk^2 \equiv 8\pi G = 8\pi/m^2_\mathrm{pl}$. We will also use the notation
$\gP\equiv\sqrt{\gP_B \gP^B}$.

The system of equations (\ref{H_dynamics})--(\ref{0i}) provides a
basis for what has been termed in the past the `separate universe
picture' for studying the evolution of perturbations on long
wavelengths. According to this picture, each point of the
long-wavelength inhomogeneous universe evolves like a separate FRW
universe. In general, this picture must be supplemented by the
constraint (\ref{0i}) which connects the separate universes
together. Only when the scalar fields are all slowly rolling can
the gradient constraint (\ref{0i}) be ignored, since in this case
it can be obtained from the Friedmann equation (\ref{friedmann})
with the $\gP^2$ term dropped and the field equation
(\ref{momentum_dynamics}) with the left-hand side set to zero.
However, we will not make any slow-roll assumptions for
the formalism we develop and hence we keep both
(\ref{friedmann}) and (\ref{0i}). All formulae in this paper are valid
regardless of slow roll.

Equations (\ref{H_dynamics})--(\ref{0i}) can be reformulated in a
convenient way to make easy contact with the well-known theory of
gauge-invariant linear perturbations, but which is fully
non-linear. In the separate universe picture deviations from
uniformity can be described as differences in the properties of
`neighbouring universes'. It thus makes sense to consider spatial
gradients in order to describe perturbations. The use of spatial gradients
was first advocated in \cite{ellis} in the context of the covariant
formalism. Here we consider two spacetime
scalars $A$ and $B$, and define the following combination of their
spatial gradients:
	\beq\label{general_gi}
	\cC_i\equiv\der_iA-\frac{\der_t A}{\der_t B}\,\der_i B\,,
	\eeq
as in \cite{gp,gp2}. We show in the appendix that under
long-wavelength changes of time
slicing $(t,\vc{x})\rightarrow(\ttt,\tilde{\vc{x}})$,
	\beq
	\cC_\ti=\delta^j{}_\ti\,\cC_j\,,
	\eeq
which means that such
combinations of spatial gradients are invariant under these
specific coordinate transformations. One such quantity that will
be very useful is \cite{gp2}
	\beq\label{g.i.var2}
	\cQ_i^A=\mathrm{e}^\ga\left(\partial_i\phi^A-\frac{\gP^A}{H}\,
	\der_i\ga\right)
	\eeq 
(the fact that $\ga$ transforms as a
scalar is also shown in the appendix). Note that when linearised
around a homogeneous background, $\cQ_i^A$ is just the gradient of
the well-known Sasaki-Mukhanov variable \cite{mfb, vantent}
	\beq\label{s-m}
	q^A_\mathrm{lin}=a(t)\left(\delta\phi^A
	+\frac{\dot{\phi}^A(t)}{N(t)H(t)}\,\Psi\right),
	\eeq
where $\Psi$ is
the perturbation in the trace of the spatial metric. Recently, the
authors of \cite{langlois-vernizzi} used similar combinations of
gradients in the context of the covariant formalism \cite{ellis}
for a single fluid
and showed that second-order gauge-invariant variables
given in the literature can be derived from them. Since they are
non-perturbative, presumably they exhibit gauge invariance to all
orders in the long-wavelength approximation.

From (\ref{H_dynamics})--(\ref{0i}) an equation for $\cQ^A_i$ can
be derived \cite{gp2}: 
	\beq\label{basic}
	{\cD_t^2}\cQ^A_i-\left(\frac{\dot{N}}{N}-NH\right){\cD_t}\cQ^A_i
	+(NH)^2\,\Omega^A{}_B\cQ^B_i=0\,, 
	\eeq 
with the ``mass matrix''
	\beq\label{omega} 
	\Omega^A{}_B \equiv \frac{V^A{}_B}{H^2}
	-\frac{2\tge}{\gk^2} R^A{}_{DCB} \frac{\gP^D}{\gP}
	\frac{\gP^C}{\gP} -\left(2-\tge\right)\delta^A{}_B
	-2\tge\left[(3+\tge)\frac{\gP^A}{\gP}\frac{\gP_B}{\gP}
	+\frac{\gP^A}{\gP}\,\tget_B+\tget^A\frac{\gP_B}{\gP}\right]\,,
	\eeq 
where $V^A{}_B \equiv G^{AC}\cD_B V_C \equiv G^{AC}(\der_B
V_C - \gG^D_{BC} V_D)$, and $R^A{}_{DCB}$ is the curvature tensor
of the field manifold. We define 
	\beq\label{sr1} 
	\tge \equiv
	\frac{\gk^2}{2}\frac{\gP^2}{H^2}\,, 
	\qquad\qquad 
	\tget^A \equiv
	-\frac{3H\gP^A+G^{AB}V_B}{H\gP}\,, 
	\qquad\qquad 
	\tgx^A \equiv -
	\frac{V^A{}_B}{H^2} \frac{\gP^B}{\gP} + 3 \tge \,
	\frac{\gP^A}{\gP} - 3 \tget^A. 
	\eeq 
In the absence of stochastic
source terms (see next section) the relations (\ref{sr1}) are
equivalent to the standard definitions of the multiple-field
slow-roll parameters \cite{vantent} (where one should read
$(1/N)\cD_t$ instead of $\cD$) by using (\ref{H_dynamics}) and
(\ref{momentum_dynamics}) and its time derivative. If a slow-roll
approximation were to be made, $\tge$ and $\tget^A$ would be first
order, while $\tgx^A$ would be second order. However, we stress
again that no slow-roll approximation is made in this paper, so no
restriction is placed on the magnitude of $\te$, $\tet^A$, and
$\tgx^A$. Equation (\ref{basic}) describes the full non-linear
dynamics on long wavelengths. Although it looks linear in
$\cQ^A_i$, its coefficients are spatially dependent functions
which depend implicitly on $\cQ^A_i$ (see \cite{gp2} for details).
It is valid for any choice of time slicing.

Equation (\ref{basic}) easily connects with linear perturbation
theory, since its linearised version is the spatial gradient of
the linear equation for $q^A_\mathrm{lin}$ \cite{vantent} in the
long-wavelength approximation. For our purposes another variable
will be more convenient than $\cQ_i^A$. We define
	\beq\label{zeta_def}
	\gz^A_i\equiv-\frac{\gk}{e^{\alpha}\sqrt{2\te}}\,\cQ^A_i\,,
	\eeq
which when linearised is just the spatial gradient of the
well-known comoving curvature perturbation $\zeta$. Just as $\cQ_i^A$, the
variable $\gz_i^A$ is invariant under choices of time slicing within the
long-wavelength approximation.\footnote{We note that we can write
$(\gP_A/\gP)\zeta^A_i=\partial_i\alpha-(\der_t \ga/\der_t \rho)\partial_i\rho$, 
where $\rho$ is the energy density of the scalar fields (here we have used 
(\ref{g.i.var2}), the definition of $\tge$ in (\ref{sr1}), (\ref{0i}), 
(\ref{friedmann}) in the form $H^2 = \gk^2 \gr/3$, (\ref{H_dynamics}), and 
(\ref{hubble})). This shows that in the
long-wavelength approximation the variable used in \cite{langlois-vernizzi} is 
identical to the component of our $\gz^A_i$ parallel to the field velocity,
i.e.\ its single-field version.}  
Expressing the long-wavelength evolution equation (\ref{basic}) in terms of
$\zeta^A_i$ we get
	\beq\label{basic zeta}
	\cD_t^2\zeta^A_i - \left(\frac{\dot{N}}{N}-2NH \left(\frac{3}{2}
	+\te+\tet^\| \right)\right)\cD_t\zeta^A_i +
	(NH)^2\,\Xi^A{}_B\zeta^B_i=0\,,
	\eeq
with
	\bea
	\Xi^A{}_B  &\equiv&  \frac{V^A{}_B}{H^2}-\frac{2\tge}{\gk^2} R^A{}_{DCB} 
	\frac{\gP^D}{\gP} \frac{\gP^C}{\gP} + \left(3\te+3\tetp+2\te^2+4\te\tetp 
	+ (\tet^\perp)^2 + \tgx^\| \right)\delta^A{}_B \nonumber \\
	&&-2\te\left((3+\te)\frac{\gP^A}{\gP}\frac{\gP_B}{\gP}
	+\frac{\gP^A}{\gP}\,\tget_B+\tget^A\frac{\gP_B}{\gP}\right)\,,
	\eea
and
	\beq\label{tetptetperp} 
	\tetp\equiv\frac{\gP^A}{\gP}\tet_A\,,
	\qquad\qquad \tet^{\perp}\equiv\left|\left(\delta^A{}_B
	-\frac{\gP^A\gP_B}{\gP^2} \right)\tet^B\right|\,,
	\qquad\qquad
	\tgx^{\parallel}\equiv
	-\frac{\gP^A}{\gP}\frac{V_{AB}}{H^2}\frac{\gP^B}{\gP} 
	+3\te-3\tetp\,. 
	\eeq
Since the coefficients appearing in (\ref{basic zeta}) are
spatially dependent, a set of constraint equations is needed to
close the system \cite{gp2}:
	\bea\label{constr_zeta_0}
	\frac{\d}{\d t}\partial_i\ga &=& -NH\te\,\partial_i\ga + H\partial_iN + NH
	\te\,\frac{\gP_A}{\gP}\zeta^A_i\,, \\
	\label{constr_zeta_1}\partial_i\ln H
	&=&\te\left(\frac{\gP_A}{\gP}\zeta^A_i-\partial_i\alpha\right)\,,\\
	\label{constr_zeta_2}\partial_i\phi^A &=& \frac{\sqrt{2\te}}{\kappa}
	\left(\frac{\gP^A}{\gP}\partial_i\ga-\zeta^A_i\right)\,,
	\eea
and
	\beq\label{constr_zeta_3}
	\cD_i\gP^A=-\frac{\sqrt{2\te}}{\kappa} 
	\left[\frac{1}{N}\cD_t\zeta^A_i + H\left((\te+\tet^\|)\delta^A{}_B
	-\te\,\frac{\gP^A}{\gP}\frac{\gP_B}{\gP}\right)\zeta^B_i 
	- H \tet^A \partial_i\alpha \right]\,.
	\eeq
With these constraint relations the spatial derivative of any
quantity of interest can be calculated in terms of $\zeta^A_i$ and its time
derivative, for a given choice of time slicing (i.e.\ the lapse function $N$). 
Equation (\ref{basic zeta}) along with the constraints
(\ref{constr_zeta_0})--(\ref{constr_zeta_3}) are the main results of this
section.

For single-field inflation an important result can be obtained
immediately, since in that case $\Xi^A{}_B$ in (\ref{basic zeta}) vanishes
identically, $\Xi^A{}_B=0$. Hence, $\zeta_i$ is seen to be
conserved in single-field inflation, as was first shown in \cite{gp,gp2}. The
authors of \cite{langlois-vernizzi} recently reached a similar
conclusion. With
a choice of time slicing which sets $\der_i\phi=0$,
$\zeta_i=\der_i\ga$ corresponds to the gradient of the integrated
expansion $\ga(t,\vc{x})=\int NH\, \d t$ of different points. So,
in single-field inflation the difference in the number of e-folds
of the separate universes is conserved. The conservation law has
been formulated in these terms in \cite{lms}.

Before closing this section we would like to make a few remarks on
the `$\delta N$ formalism', which has been used in other, more
recent work (see e.g.\ \cite{deltaN})\footnote{Note that in these other 
works $N(t,\mathbf{x})$
is the local expansion, not the lapse function. In our notation it
would be the `$\delta \alpha$ formalism'.} as an alternative for
studying non-linear perturbations.  In this formalism the
perturbed expansion $\alpha(t,\mathbf{x})$ on comoving time slices
is related to the perturbations of the scalar fields
$\delta\phi^A(\mathbf{x})$ on an initial flat time slice where the
perturbations are supposed to have been generated after horizon
crossing. In our notation
$\delta\alpha(t,\mathbf{x})=\partial^{-2}\partial_i\left(
\left(\gP_A/\gP\right)\zeta^A_i\right)$
since $\left(\gP_A/\gP\right)\zeta^A_i=\partial_i\alpha$ on
comoving time slices. All calculations in the $\delta N$ formalism
are based on the knowledge of $\alpha(t,\phi^A(\mathbf{x}))$,
i.e.\ the local expansion expressed in terms of the initial fields
at each spatial point. All the initial scalar field values which appear in 
this solution are then made spatially dependent. Practical calculations 
require an accurate multi-dimensional solution for the expansion
$\alpha(t,\phi^A)$ of the homogeneous universe, from which all its field
derivatives can be derived. Furthermore, this entails the approximation that
the momentum dependence is completely ignored, since in general
$\ga(t,\gf^A,\gP^A)$ is a function of both the fields and the momenta. The
conditions for such an approach to be valid can be seen from 
(\ref{H_dynamics})--(\ref{0i}). Since the evolution equations on long 
wavelengths are the same as those of a homogeneous cosmology valid locally,
any analytic solution of a homogeneous universe with the initial
conditions made spatially varying will provide a solution
describing the evolution of the true inhomogeneous spacetime. The
initial conditions must satisfy both of the constraints (\ref{friedmann}) and
(\ref{0i}). In \cite{deltaN} equation
(\ref{0i}) is ignored, which is the same approximation as
ignoring the momenta (see the discussion below (\ref{0i})). 
These conditions restrict the type of
multiple-field models that can be investigated. In contrast,
the formalism presented here is applicable in a practical way to
any expansion history, slow-roll or otherwise, and it is tractable
numerically when analytic solutions cannot be obtained in full.

\section{Stochastic Equations}
\label{stochsec}

In the previous section we derived equations for the
long-wavelength evolution of the variables $\gz^A_i$ defined in
(\ref{zeta_def}). These are (\ref{basic zeta}) along with the
constraints (\ref{constr_zeta_0})--(\ref{constr_zeta_3}). In order to
solve this long-wavelength system initial conditions must be
provided. In inflation these originate in the short-wavelength
quantum regime, before the fluctuations in the metric and matter cross 
the horizon, so we have to incorporate these sub-horizon effects in our 
long-wavelength equations. We achieve this in the following way. Once
super-horizon, the quantum fluctuations can be considered as
classical stochastic quantities. This is the stochastic picture
for the generation of inflationary perturbations
\cite{stoch}. Linear theory is used to describe
the perturbations until shortly after horizon crossing. These are then
used in stochastic source terms as the initial conditions for the full 
non-linear long-wavelength system. A smoothing window function is used
to separate short and long wavelengths as in \cite{gp2}. However,
improving on \cite{gp2}, we start from a first-order system, in
order to correctly reproduce the velocity modes as well.
We also note that in many references on stochastic inflation,
metric fluctuations are not consistently included. They {\em are}
included in our formalism.

For the definition of the source terms it is convenient to fix the 
remaining gauge freedom ($N$) by choosing $\gt=\ln (aH)$ as the time 
variable.\footnote{This is a useful time variable as long as it is 
monotonic, a condition that is invalid only around reheating. Since
the source terms have disappeared by then, a different choice can be
made near the end of inflation.} One advantage of this gauge choice is 
that horizon crossing of modes ($k=aH$) happens simultaneously throughout 
the whole universe. Other advantages will become clear below. For this
choice we have
	\beq\label{gauge} 
	NH=\frac{1}{1-\te}\,,
	\eeq
and our basic equations (\ref{basic zeta}), 
(\ref{constr_zeta_1})--(\ref{constr_zeta_3}) read (note that
(\ref{constr_zeta_0}) is no longer an independent equation)
	\beq\label{zeta special}
	\cD_\gt^2\gz^A_i+\frac{3-2\te+2\tetp-3\te^2-4\te\tetp}{(1-\te)^2}\,
	\cD_\gt\gz^A_i+\frac{1}{(1-\te)^2}\,\Xi^A{}_B\,\gz^B_i=0\,, 
	\eeq
with
	\bea\label{constr_zeta_special_1}
	\der_i\ga & = & -\partial_i(\ln H) \: = \:
	-\frac{\te}{1-\te}\,\frac{\gP_A}{\gP}\,\gz^A_i\,,
	\\
	\partial_i\phi^A & = & -\frac{\sqrt{2\te}}{\gk}\left(\delta^A{}_B
	+\frac{\te}{1-\te} \frac{\gP^A}{\gP}\frac{\gP_B}{\gP}\right)\gz^B_i\,,
	\eea
and
	\beq\label{constr_zeta_special_3}
	\cD_i\gP^A=-\frac{\sqrt{2\te}}{\gk}H\left[(1-\te){\cD_\gt}\gz^A_i
	+\left((\te+\tetp)\delta^A{}_B-\te\,\frac{\gP^A}{\gP}\frac{\gP_B}{\gP}
	+\frac{\te}{1-\te}\,\tet^A\frac{\gP_B}{\gP}\right)\gz^B_i\right]\,.
	\eeq

Defining the velocity $\gth_i^A$ we express the dynamical equation
(\ref{zeta special}) as 
	\beq\label{basic_zeta_split} 
	\left\{ \begin{array}{l}
	\displaystyle\cD_\gt\gz^A_i-\theta^A_i=0\\
	\displaystyle \cD_\gt\theta^A_i
	+\frac{3-2\te+2\tetp-3\te^2-4\te\tetp}{(1-\te)^2}\,\theta^A_i
	+\frac{1}{(1-\te)^2}\,\Xi^A{}_B\,\gz^B_i=0
	\end{array}\right.
	\eeq
We then add to the right-hand side of these equations
stochastic source terms to emulate the continuous influx of
short-wavelength modes when they cross the horizon, in this way
setting up the proper initial conditions for the long-wavelength
system. The derivation of these source terms goes as follows (for
more details see \cite{gp2}). We start from the exact (i.e.\ no
long-wavelength approximation) equations for the linear theory,
written as two first-order differential equations like
(\ref{basic_zeta_split}). We define smoothed long-wavelength
variables using a window function $\cW$. In Fourier space this
means $\gz^A_\mathrm{lin, lw}(k) = \cW(k) \gz^A_\mathrm{lin}(k)$,
and an identical expression for $\gth^A$. Rewriting the exact
linear equations in terms of the smoothed variables
$\gz^A_\mathrm{lin, lw}$, one is left with terms that depend on
the non-smoothed variables $\gz^A_\mathrm{lin}$. These terms
together form the source, and are put on the right-hand side of
the equations, while the other terms are on the left-hand side. At
the linearised level these equations are still exact. Next, in
order to go to the non-linear case one uses the fact that the
left-hand side is exactly the linearised version of the
long-wavelength system (\ref{basic_zeta_split}). It is then
postulated that the right-hand side can be taken as the source
term for the non-linear long-wavelength equations, if one replaces
all background quantities with their fully inhomogeneous versions.
The final result is:
	\beq\label{st_main_1}
	\left\{
	\begin{array}{l}
	\displaystyle \cD_\gt\gz^A_i-\theta^A_i=\cS^A_i\\
	\displaystyle \cD_\gt\theta^A_i
	+\frac{3-2\te+2\tetp-3\te^2-4\te\tetp}{(1-\te)^2}\,\theta^A_i
	+\frac{1}{(1-\te)^2}\,\Xi^A{}_B\,\gz^B_i=\cJ^A_i
	\end{array}\right.
	\eeq
with the source terms $\cS^A_i$ and $\cJ^A_i$ given by
	\beq\label{sources}
	\cS^A_i \equiv \dk \, \dot{\cW}(k) \,
	\zeta_\mathrm{lin}^A(\vc{k},\vc{x}) \, \mathrm{i}k_i
	\,\mathrm{e}^{\mathrm{i} \vc{k}\cdot\vc{x}} +\mathrm{c.c.}\,, \qquad
	\cJ^A_i \equiv \dk \, \dot{\cW}(k) \,
	\theta_\mathrm{lin}^A(\vc{k},\vc{x}) \, \mathrm{i}k_i \,
	\mathrm{e}^{\mathrm{i} \vc{k}\cdot\vc{x}} +\mathrm{c.c.}\,,
	\eeq
where c.c.\ denotes the complex conjugate and $\gz^A_\mathrm{lin}$ and
$\gth^A_\mathrm{lin}$ are the full, non-smoothed solutions from
linear perturbation theory, that is, incorporating
short-wavelength information. The fact that they depend on
$\vc{x}$ as well as on $\vc{k}$ represents the fact that all
background quantities in these solutions should be made inhomogeneous.

The following relations hold:
	\beq\label{gzqrel}
	\gz^A_\mathrm{lin} = \frac{-\gk}{a\sqrt{2\te}}\, q^A_\mathrm{lin}\,,
	\qquad\qquad
	\theta_\mathrm{lin}^A = \cD_\gt  \gz^A_\mathrm{lin}\,,
	\qquad\qquad
	q^A_\mathrm{lin}(\vc{k}) = Q^A_{\mathrm{lin}\,B}(k) \ga^B(\vc{k}),
	\eeq
where $q^A_\mathrm{lin}$ is the solution
from linear theory for the Sasaki-Mukhanov variable (\ref{s-m}).
In other words, $Q^A_{\mathrm{lin}\,B}(k)$ is the solution of \cite{vantent}
	\beq\label{linearQ}
	{\cD_\gt^2}Q^A_{\mathrm{lin}\,B}
	+\frac{1-2\tge-\tge^2-2\tge\tget^\parallel}{(1-\tge)^2} \,
	{\cD_\gt}Q^A_{\mathrm{lin}\,B}
	+\frac{\Omega^A{}_C}{(1-\tge)^2} \, Q^C_{\mathrm{lin}\,B}
	+ \frac{k^2}{(aH)^2 (1-\tge)^2} \, Q^A_{\mathrm{lin}\,B} = 0\,,
	\eeq
where all coefficients take their homogeneous background values and with 
initial conditions deep within the horizon
$Q^A_{\mathrm{lin}\,B}(k) = U^A{}_B/\sqrt{2k}$ and
$\cD_\gt Q^A_{\mathrm{lin}\,B}(k) = 1/[aH(1-\tge)] \, \mathrm{i}\sqrt{k/2}
\, U^A{}_B$, where $U^A{}_B$ is a physically irrelevant unitary matrix.
Then the sources can be written as
	\bea
	\cS_i^A & = & -\frac{\gk}{a\sqrt{2\tge}} \dk \, \dot{\cW}(k)
	Q^A_{\mathrm{lin}\,B}(k) \ga^B(\vc{k}) \,
	\mathrm{i} k_i \mathrm{e}^{\mathrm{i} \vc{k}\cdot\vc{x}}
	+ \mathrm{c.c.}, \non\\
	\cJ_i^A & = & -\frac{\gk}{a\sqrt{2\tge}} \dk \, \dot{\cW}(k)
	\left[ \cD_\gt Q^A_{\mathrm{lin}\,B}(k)
	-\frac{1+\tge+\tget^\parallel}{1-\tge}\,Q^A_{\mathrm{lin}\,B}(k) \right]
	\ga^B(\vc{k}) \,
	\mathrm{i} k_i \mathrm{e}^{\mathrm{i} \vc{k}\cdot\vc{x}}
	+ \mathrm{c.c.}
	\label{sourcesQ}
	\eea
Following \cite{gp2} we have introduced a set of complex Gaussian
stochastic quantities $\ga^A(\vc{k})$ to replace the quantum creation
and annihilation operators, satisfying
	\beq \label{correlation}
	\langle\alpha^A(\vc{k})\alpha_B^{*}(\vc{k}')\rangle
	=\delta^{A}{}_B\,\delta^3(\vc{k}-\vc{k}')\,,
	\qquad\qquad
	\langle\alpha^A(\vc{k})\alpha_{B}(\vc{k}')\rangle=0,
	\eeq
where $\langle\ldots\rangle$ denotes an ensemble average. This is why
the source terms are called stochastic.
Because $\gz^A_i$ and $\gth^A_i$ now represent
smoothed, long-wavelength variables, it is clear that they have to
be zero at early times when all the modes are sub-horizon:
	\beq\label{initial}
	\lim_{\gt\rightarrow-\infty} \gz^A_i = 0, \qquad\qquad
	\lim_{\gt\rightarrow-\infty} \gth^A_i = 0.
	\eeq
The appropriate short-wavelength initial conditions are then introduced 
into the system later via the stochastic source terms.

The linear perturbation equation (\ref{linearQ}) can either be solved exactly 
numerically to obtain $Q^A_{\mathrm{lin}\,B}$, or analytically within the 
slow-roll approximation. The latter was done in \cite{vantent}. From that 
analytic slow-roll solution one finds that 
$\cD_\gt Q^A_{\mathrm{lin}\,B} = \tD^A{}_C Q^C_{\mathrm{lin}\,B}$,
with $\tD^A{}_C$ a matrix containing slow-roll parameters (given explicitly in
\cite{vantent}), and 
$Q^A_{\mathrm{lin}\,B} = c/(2 k^{3/2} R)\,\gd^A_B$ + first-order slow-roll 
terms (omitting physically irrelevant overall unitary factors). 
Even though in this paper we make no use of the slow-roll approximation
whatsoever, for applications of the formalism it is useful to observe that 
this means that, in our gauge, non-linear corrections to $Q_\mathrm{lin}$ 
are higher order in slow roll and can be neglected in a leading-order treatment. 
(This reflects the fact that $q$, not $\gz$, is the proper well-behaved 
quantity to use on short wavelengths.) Hence in such a treatment the only 
quantities that have to be made 
inhomogeneous in the source terms (\ref{sourcesQ}) are the matrix $\tD^A{}_C$ 
and the $1/(a\sqrt{2\tge})$ prefactors. This point is explicitly
worked out in \cite{sf,mf}.

We can take the Fourier transform of the window function $\cW(k)$
to be a Gaussian, 
	\beq 
	\cW(k) \equiv \mathrm{e}^{-k^2 R^2/2} \,,
	\qquad\qquad 
	R \equiv \frac{c}{aH} = c\, \mathrm{e}^{-\gt}, 
	\eeq
with $R$ the smoothing length above which the system is considered
to be `long wavelength'. We take it to be a small multiple of the
comoving Hubble radius, since the latter is the natural scale that
separates short and long wavelengths in inflation;
$c\approx\,$3--5 will do for our purposes. Since the smoothing
length is decreasing quasi-exponentially during inflation, more
and more modes will populate the long-wavelength system. For our
choice of window function, $\dot{\cW}(k)=
(kR)^2\cW(k)$, where a dot denotes a time derivative with respect
to $\gt=\ln(aH)$. If the smoothing length $R$ were constant,
$\cS^A_i$ and $\cJ^A_i$ would be zero. Note that with our choice
of time the smoothing length remains uniform across space even in
the presence of perturbations. Had we not made this choice any
mode would enter the long-wavelength system at different times for
different points in space, which would complicate matters. The
system of equations (\ref{st_main_1}) along with the constraints
(\ref{constr_zeta_special_1})--(\ref{constr_zeta_special_3}) (where 
$\cD_\gt \gz^A_i$ is replaced by $\gth^A_i$) forms a consistent and
self-contained system of non-linear stochastic equations and is
the main result of this paper. When linearised it is exact and
reproduces the well-known gauge-invariant linear perturbation
theory \cite{mfb}.

At first sight it might seem that any results obtained from
solving (\ref{st_main_1}) will depend on the {\it ad hoc} choice
for the window function $\cW$. It turns out that the exact form of
$\cW$ is for the most part irrelevant. At linear order there is no
dependence of the final results on the exact form of $\cW$, by construction; 
any properly normalised function with $\cW(k)\rightarrow 1$ for scales
sufficiently larger than the horizon will produce the same final
answer. Beyond linear order there is some dependence on the choice of
$\cW$, but this dependence is limited to terms that involve only non-linear
effects around horizon crossing, which turn out to be small and hence 
uninteresting observationally. The non-Gaussianity in single-field inflation
is an example of this, see \cite{sf}, but even here only the limit with all
three momenta of the same order is affected by the choice of window function. 
On the contrary, any effects which 
involve super-horizon evolution are independent of the functional
form of $\cW$ for scales sufficiently larger than the horizon. We
discuss these issues in greater detail in \cite{mf,mf2}, where we also
show that such super-horizon evolution effects can lead to potentially
observably large non-Gaussianity.

\section{Field Basis}
\label{basissec}

So far we have used a field component notation, with indices $A,B$, etc.\
labeling the fields. However, in actual analytic multiple-field calculations it
is more convenient to work with a certain explicit basis on the field manifold,
which allows us to clearly distinguish effectively single-field effects from
truly multiple-field ones, to work again with normal time derivatives instead 
of covariant ones, and which is also a necessary ingredient for quantisation.
This basis was first defined in \cite{vantent1}, and is also described in
\cite{vantent}.

The first basis vector $e_1^A$ is the direction of the field
velocity. Next, $e_2^A$ is defined as the direction of that part
of the field acceleration that is perpendicular to $e_1^A$. Hence,
	\beq 
	e_1^A \equiv \frac{\gP^A}{\gP}, 
	\qquad\qquad 
	e_2^A \equiv \frac{\cD_t \gP^A - e_1^A e^{}_{1\,B} \cD_t \gP^B} 
	{|\cD_t \gP^A - e_1^A e^{}_{1\,B} \cD_t \gP^B|}. 
	\eeq 
One continues this
orthogonalisation process with higher derivatives until a complete
basis is found. Explicitly,\footnote{Again, as in the definition
of the slow-roll parameters, using (\ref{momentum_dynamics}) and
its time derivatives one can rewrite all time derivatives of
$\gP^A$ in terms of $\gP^A$ itself and field derivatives of the
potential, and define the basis vectors in that way. However, once
the stochastic source terms are included, equation
(\ref{momentum_dynamics}) changes subtly and the two ways of
defining the basis vectors (except for $e_1^A$) are no longer
equivalent. At that point it is actually the second way that is
our real definition, but since those equations are rather awkward
and have to be written down for each value of $m$ separately, we
give here the compact definition in terms of time derivatives of
$\gP^A$.} 
	\beq\label{basis} 
	e_m^A \equiv \frac{(P_{m-1}^\perp)^A{}_B \lh\frac{1}{N}\cD_t\rh^{m-1}\gP^B}
	{|(P_{m-1}^\perp)^A{}_B \lh\frac{1}{N}\cD_t\rh^{m-1} \gP^B|},
	\qquad\qquad 
	(P_m)^A{}_B \equiv e^A_m e^{}_{m\,B}, 
	\qquad\qquad
	(P_m^\perp)^A{}_B \equiv \gd^A_B - \sum_{q=1}^m (P_q)^A{}_B. 
	\eeq
Here the projection operators $P_m$ project on the $e_m$, while
the $P_m^\perp$ project on the subspace that is perpendicular to
$e_1, \ldots, e_m$, and we define $(P_0^\perp)^A{}_B \equiv
\gd^A_B$. The basis is orthonormal: $e_m^A e^{}_{n\,A} =
\gd_{mn}$. Now one can take components of vectors in this basis
and we define, for example for $\gz_i^A$, 
	\beq\label{Lmdef}
	\gz_i^m \equiv e_{m\,A} \gz_i^A. 
	\eeq 
Note that, unlike for the
index $A$, there is no difference between upper and lower indices
for the $m$. The slow-roll parameters $\tget^\parallel$ and
$\tget^\perp$ given in (\ref{tetptetperp}) are now simply
$\tget^\parallel \equiv e_1^A \tget_A$ and $\tget^\perp \equiv
e_2^A \tget_A$ (by construction there are no other components of
$\tget^A$) and equivalently $\tgx^\parallel \equiv e_1^A \tgx_A$.
The definitions (\ref{tetptetperp}) can be rewritten as 
(including $\tge$ for completeness' sake)
	\beq\label{SRrewr} 
	\tge = \frac{\gk^2 \gP^2}{2 H^2}, 
	\qquad 
	\tget^\parallel = -3 - \frac{e_1^A V_A}{H \gP},
	\qquad 
	(\tget^\perp)^2 = \frac{V^A V_A - (e_1^A V_A)^2}{H^2 \gP^2}, 
	\qquad 
	\tgx^\parallel = 3 \tge - 3 \tget^\parallel 
	- \frac{e_1^A V_{AB} e_1^B}{H^2}.
	\eeq
Finally we define for later use 
	\beq 
	Z_{mn} \equiv \frac{1}{NH} \, e_{m\,A} \cD_t e_n^A. 
	\eeq 
It is antisymmetric and only non-zero
just above and below the diagonal. If a slow-roll approximation
were to be made, it would be first order in slow roll. Its
explicit form in terms of slow-roll parameters can be found in
\cite{vantent}; up to $m,n=3$ the only non-zero terms are $Z_{12}
= - Z_{21} = -\tget^\perp$ and $Z_{23} = - Z_{32} =
-\tgx_3/\tget^\perp$.

The equation of motion (\ref{zeta special}) for $\gz_i^A$ can now
be rewritten as an equation of motion for $\gz_i^m$: 
	\beq
	\ddot{\gz}_i^m + \lh \frac{(3 - 2\tge + 2\tget^\parallel - 3\tge^2
	- 4\tge\tget^\parallel)\gd_{mn}}{(1-\tge)^2} +
	\frac{2Z_{mn}}{1-\tge} \rh \dot{\gz}_i^n +
	\frac{\gX_{mn}}{(1-\tge)^2} \, \gz_i^n = 0, 
	\eeq 
where $\dot{\gz}_i^m \equiv \frac{\der}{\der\gt} (\gz_i^m) =
\frac{\der}{\der\gt} (e^{\;}_{m\,A} \gz_i^A)$, with $\gt$ the time
variable defined in (\ref{gauge}), and with 
	\bea 
	\gX_{mn} & \equiv & \frac{V_{mn}}{H^2} - \frac{2\tge}{\gk^2} R_{m11n} +
	(1-\tge)\dot{Z}_{mn} + Z_{mp}Z_{pn} 
	+ \lh 3-2\tge+2\tget^\parallel-\tge^2-2\tge\tget^\parallel \rh
	\frac{Z_{mn}}{1-\tge} \\
	&& + \lh 3\tge+3\tget^\parallel+2\tge^2+4\tge\tget^\parallel
	+(\tget^\perp)^2+\tgx^\parallel \rh \gd_{mn}
	- 2\tge \lh (3+\tge+2\tget^\parallel) \gd_{m1}\gd_{n1} 
	+ \tget^\perp (\gd_{m1}\gd_{n2}+\gd_{m2}\gd_{n1}) \rh, \non
	\eea
where $R_{m11n} \equiv e_m^A R_{ABCD} e_1^B e_1^C e_n^D$ and
$V_{mn} \equiv e_m^A V_{AB} e_n^B$. The matrix $\gX_{mn}$ is not just
$e_m^A \gX_{AB} e_n^B$, but contains additional $Z_{mn}$ terms.
Note that $\gX_{11} = 0$ identically, which means that in the single-field case
$\gX$ is zero, and hence $\gz_i$ conserved on long wavelengths, as we
mentioned before.
In a procedure completely identical to the one described in the previous
section, we can then split the second-order differential equation into two
first-order equations and add stochastic source terms to find the analogue of
(\ref{st_main_1}):
	\beq\label{st_main_basis}
	\left\{ \begin{array}{l}
	\displaystyle
	\dot{\gz}_i^m - \gth_i^m = \cS_i^m\\
	\displaystyle
	\dot{\gth}_i^m + \lh \frac{(3 - 2\tge + 2\tget^\parallel - 3\tge^2
	- 4\tge\tget^\parallel)\gd_{mn}}{(1-\tge)^2} + \frac{2Z_{mn}}{1-\tge} \rh
	\gth_i^n + \frac{\gX_{mn}}{(1-\tge)^2} \, \gz_i^n = \cJ_i^m
	\end{array} \right.
	\eeq
where the source terms $\cS_i^m$ and $\cJ_i^m$ are defined analogously to
(\ref{sources}) with $\gz^m_\mathrm{lin}$ and $\gth^m_\mathrm{lin}$ instead of
$\gz^A_\mathrm{lin}$ and $\gth^A_\mathrm{lin}$. It is important to realise that
for a velocity, like $\gth_i^m$, the relation between $\gth_i^m$ and $\gth_i^A$
is not simply a contraction with the basis vector, since the time derivative of
the basis vector has to be taken into account as well.
They are related by $\gth_i^m = e_{m\,A} \gth_i^A - NH Z_{mn} \gz_i^n$.

Of course (\ref{st_main_basis}) has to be supplemented with the constraints
(\ref{constr_zeta_special_1})--(\ref{constr_zeta_special_3}). From these 
constraints and the
definitions (\ref{SRrewr}) we can derive the following expressions for the
spatial derivatives of $a$, $H$ and the slow-roll parameters:
	\bea
	\der_i \ln a & = & - \der_i \ln H = - \frac{\tge}{1-\tge} \, \gz_i^1,
	\non\\
	\der_i \tge & = & -2 \tge \lh (1-\tge) \gth_i^1 +
	\frac{\tge+\tget^\parallel}{1-\tge} \, \gz_i^1 - \tget^\perp \gz_i^2 \rh,
	\non\\
	\der_i \tget^\parallel & = & (1-\tge) \lh (3+\tget^\parallel) \gth_i^1
	- \tget^\perp \gth_i^2 \rh
	- \frac{1}{1-\tge} \lh (3+\tget^\parallel)(\tge-\tget^\parallel)
	+ (\tget^\perp)^2 - \frac{V_{11}}{H^2} \rh \gz_i^1
	- (3+\tge+2\tget^\parallel)\tget^\perp \gz_i^2
	\non\\
	&& + \tgx_3 \, \gz_i^3 + \sum_{n\geq 2}\frac{V_{1n}}{H^2}\,\gz_i^n,
	\label{deriSR}\\
	\der_i \tget^\perp & = & (1-\tge) \lh \tget^\perp \gth_i^1
	+ (3+\tget^\parallel) \gth_i^2 \rh
	+ \frac{1}{1-\tge} \lh (3-\tge+2\tget^\parallel)\tget^\perp
	+ \frac{V_{12}}{H^2} \rh \gz_i^1
	+ \lh (3+\tget^\parallel)(\tge+\tget^\parallel) - (\tget^\perp)^2 \rh 
	\gz_i^2 \non\\
	&& - (3+\tget^\parallel)\frac{\tgx_3}{\tget^\perp}\,\gz_i^3
	+ \sum_{n\geq 2}\frac{V_{2n}}{H^2} \, \gz_i^n.
	\non
	\eea
When no stochastic source terms
are present, or when all slow-roll parameters take their homogeneous background
values, the relation $V_{1n}/H^2 = (3\tge-3\tget^\parallel-\tgx^\parallel)
\gd_{n1} - (3\tget^\perp+\tgx_2)\gd_{n2} - \tgx_3 \gd_{n3}$ can be used. This
relation is derived by taking the time derivative of (\ref{momentum_dynamics})
and using the definitions of the slow-roll parameters.
Let us reiterate that all these expressions are not slow-roll approximated, even
though they contain slow-roll parameters. The equations in this section will
be the basis of more detailed multiple-field calculations in \cite{mf}.

\section{Applying the Formalism}
\label{analnumsec}

The closed system of non-linear equations for $\zeta^A_i$ is, in principle,
amenable to direct numerical solution on a multi-dimensional spatial grid.
However, the inhomogeneous
coefficients in (\ref{st_main_1}) have an implicit dependence on $\zeta^A_i$,
meaning that $a(t,\vc{x})$, $\phi^A(t,\vc{x})$ and $\gP^A(t,\vc{x})$
need to be reconstructed from the constraints
(\ref{constr_zeta_special_1})--(\ref{constr_zeta_special_3}) at each timestep.  
Now in the long-wavelength approximation and in the absence of noise,
the $\zeta^A_i$ equation plus the constraints are
equivalent to the separate universe equations (\ref{H_dynamics})--(\ref{0i})
from which they were derived, that is,
the original evolution equations for $a(t,\vc{x})$, $\phi^A(t,\vc{x})$ and
$\gP^A(t,\vc{x})$.  Numerically,
it turns out to be simplest to solve the $\zeta^A_i$ equations
by supplementing them with the separate universe equations, so that the
inhomogeneous coefficients depending on $a$, $\phi^A$ and $\gP^A$
are given explicitly, rather than being reconstructed from $\zeta_i^A$.
(The redundancy entailed in this
can be viewed positively, since it gives a check on the accuracy of the
finite difference scheme through the relations (\ref{g.i.var2}) and
(\ref{zeta_def}).)

In order to add the stochastic noise on the right-hand side of
(\ref{st_main_1}), we evaluate the mode functions of the
$k$-dependent linear perturbation equations to high accuracy,
beginning with flat-space initial conditions well inside the
horizon (no slow-roll approximation is made). As in other
stochastic approaches, a realisation of the linear perturbation
solution $q_{\rm lin}^A(\vc{k},t)$ is then constructed by
multiplying random numbers satisfying (\ref{correlation}) for each
$\vc{k}$ on the numerical grid by the corresponding mode function
for $k = |\vc{k}|$. At a finely-spaced set of times, the linear
perturbations $q_{\rm lin}^A(\vc{k},t)$ are convolved with the
window function $\cW(k)$ to determine the necessary
long-wavelength correction $\delta\zeta^A_i(\vc{x},t)\delta t$ in
real space, where $\delta t$ is the timestep.  Since any change
$\delta\zeta^A_i$ necessarily entails changes to $a, \phi^A,
\gP^A$,  the corrections $\delta a, \delta\phi^A, \delta\gP^A$
also need to be determined from the constraints
(\ref{constr_zeta_special_1})--(\ref{constr_zeta_special_3}).  The
procedure then iterates through successive non-linear evolution
and stochastic noise updates. Results from the successful
numerical implementation of these equations in single- and
multiple-field inflation contexts will be presented elsewhere.

Up to this point, the only approximations that have been made are
the basic assumptions of the method: the long-wavelength
approximation and the construction of the stochastic source terms
for the non-linear equations. The equations are otherwise
completely general, that is, non-linear and no slow-roll
approximation has been made. As discussed above,
with a numerical treatment we can proceed directly with these equations.
However, when we want to use the formalism to make explicit {\em analytic}
calculations, we need to make two expansions: a slow-roll expansion, and an
expansion in perturbation orders. Of course in practice these expansions have to
be cut off at a certain order, which means that two additional approximations
are introduced.
We will not discuss the slow-roll expansion here, but describe in general terms
how a perturbative expansion can be set up. We find that compared to standard
perturbative approaches \cite{2nd_order} the equations are much simpler to
derive and use. Detailed applications are presented in separate publications
\cite{sf,mf,mf2}.

The general form of the system of non-linear stochastic equations
with the constraints is 
	\beq 
	\cD_\gt v_i+A(u)v_i=\cG_i(u),
	\qquad\qquad 
	\der_iu=C(u)v_i\,. 
	\eeq 
We use the vectors $v_i\equiv \left(\zeta^A_i,\,\theta^A_i\right)^T$ and
$u\equiv\left(\ga,\,\phi^A,\,\gP^A\right)^T$ and $\cG_i \equiv
\left(\cS_i^A,\cJ_i^A\right)^T$ (where $T$ denotes the transpose)
to simplify the notation. The quantities $A$ and $C$ are matrices,
functions of the components of $u$, which can be read off from
(\ref{st_main_1}) and
(\ref{constr_zeta_special_1})--(\ref{constr_zeta_special_3}). Now
everything can be expanded as 
	\beq 
	v_i = v_i^{(1)}+v_i^{(2)}+\ldots, 
	\qquad\qquad 
	u = u^{(0)}+u^{(1)}+u^{(2)}+\ldots 
	\eeq 
Note that $v_i$ does not have a zeroth-order spatially homogeneous 
background part. At first order we have 
	\beq\label{lin_dyn} 
	\cD_\gt v^{(1)}_i+A^{(0)}v^{(1)}_i=\cG^{(1)}_i, 
	\qquad\qquad 
	\der_i u^{(1)}=C^{(0)}v^{(1)}_i\,. 
	\eeq 
The first equation is linear in
$v_i^{(1)}$ and can be solved since the background solution and
the linear source term $\cG_i^{(1)}$ are assumed to be known. The
procedure is equivalent to standard gauge-invariant linear
perturbation theory. With this solution the constraint can then be
solved for $u^{(1)}$: 
	\beq 
	u^{(1)}=C^{(0)}\der^{-2}\der^i v^{(1)}_i. 
	\eeq

It is now straightforward to obtain the second-order perturbation
equations:
	\beq\label{2_dyn}
	\cD_\gt v^{(2)}_i+A^{(0)}v^{(2)}_i=
	\cG^{(2)}_i-A^{(1)}v_i^{(1)},
	\qquad\qquad
	\der_i u^{(2)}=C^{(0)}v^{(2)}_i + C^{(1)} v^{(1)}_i.
	\eeq
Here $A^{(1)}$ is given by
	\beq
	A^{(1)}=\lh\frac{\der A}{\der u}\rh^{(0)} u^{(1)},
	\eeq
and similarly for $C^{(1)}$ and $\cG_i^{(2)}$. We see that the equation of
motion for $v_i^{(2)}$ is again linear, with the right-hand side
known from the solution at first order, and it can be solved. Again,
given the solution for $v_i^{(2)}$ we can then solve the constraint for
$u^{(2)}$:
	\beq
	u^{(2)}=C^{(0)}\der^{-2}\der^i v^{(2)}_i
	+\der^{-2}\der^i\left(C^{(1)}v^{(1)}_i\right).
	\eeq
Knowledge of $v_i^{(2)}$ and $u^{(2)}$ gives $v_i^{(3)}$ which in turn
determines $u^{(3)}$, etc. Continuing in
this fashion one can in principle obtain all quantities of
interest to any desired order. Even though the discussion in this
section might seem rather abstract, we show in \cite{sf,mf,mf2} that this
approach works well in explicit
calculations. The expressions on the right-hand side of
(\ref{2_dyn}) are relatively simple to derive, which makes the
calculations tractable.

\section{Summary}

In this paper we described a scheme for studying the non-linear
evolution of inhomogeneities during multiple-field inflation,
extending ideas presented in \cite{gp2}. Here we summarise the
main ingredients of the formalism. {\em (i)} We advocate the use
of variables like $\zeta^A_i$, defined in (\ref{zeta_def}), using
spatial gradients to describe deviations from a homogeneous
spacetime. These variables are invariant under long-wavelength
changes of time slicing and include both matter and metric
perturbations. {\em (ii)} We derived a set of exact, non-linear,
long-wavelength evolution equations for $\gz^A_i$ (\ref{basic
zeta}). To close the system, a set of constraints
(\ref{constr_zeta_0})--(\ref{constr_zeta_3}) determines all local
quantities appearing in the coefficients of (\ref{basic zeta}). An
obvious consequence of these equations is that for single-field
inflation the variable $\zeta^A_i$ is exactly conserved. {\em
(iii)} In order to incorporate the influence of short-wavelength
quantum modes, stochastic sources are added on the right-hand side
of the evolution equations, leading to (\ref{st_main_1}),
(\ref{sources}). When linearised, the equations are exactly
equivalent to linear, gauge-invariant perturbation theory. We also
introduced the field basis of \cite{vantent} and rewrote our
equations in terms of it, which is very convenient for actual
multiple-field calculations. The full stochastic system is derived
without any slow-roll approximations and is well-suited for
numerical simulations. If slow roll is assumed, analytic results
can be obtained via perturbative expansions. The merit of
perturbing this set of non-linear equations for the variables
$\zeta^A_i$ is a significant simplification compared to the
standard approach of perturbing the Einstein equations from the
outset. This paper concentrated on describing the method, while
quantitative calculations using this formalism are given in
related publications \cite{sf,mf,mf2}.

\section*{Acknowledgements}
This research is supported by PPARC grant PP/C501676/1.

\appendix
\section{Changing the time slicing on long wavelengths}
\label{appendix}

In this appendix we demonstrate that (\ref{general_gi}) is invariant
under changes of time slicing. Again, we restrict
attention to metrics like (\ref{adm}) which do not mix temporal and
spatial intervals. The transformations we
consider are defined by going to a new time slicing $t
\rightarrow \tilde{t}(t,\vc{x})$ and then fixing the spatial
coordinates by demanding that $g_{\tilde{t}\ti}=0$, that is, making
the coordinate lines $\tx^\tj=\mathrm{const.}$ normal to
the $\ttt=\mathrm{const.}$ hypersurfaces. We show, following \cite{sb}, that,
up to second order in spatial gradients, the spatial coordinates can be
taken to transform trivially.

Suppose we are interested in a patch of spacetime which at early
enough times is homogeneous, in accordance with the initial
conditions, that is, the comoving horizon is initially taken to be
larger than the spatial extent of the patch. Perturbations are
introduced later as wavelengths shorter than the spatial region of
interest exit the horizon. While the patch is homogeneous, there
is a preferred time slicing which respects the spatial symmetries,
and spatial coordinates are chosen on these initial time slices.
After perturbations are introduced there is no preferred choice of
time slicing. Hence one should consider the consequences of
changes $t\rightarrow \tilde{t}(t,\vc{x})$, for example between
uniform field and uniform expansion time slices. To define the new
spatial coordinate lines we demand that the curves
$\tilde{x}^\tj=\mathrm{const.}$ be the integral curves of $w^\gm$
with 
	\beq\label{normal} 
	w^\gm=\frac{\d x^\gm}{\d s} =\der^\gm \tilde{t}\,. 
	\eeq 
Roman indices run from 1 to 3 and Greek from 0
to 3. The vector $w^\gm$ is normal to the
$\tilde{t}=\mathrm{const.}$ hypersurfaces. For a displacement along
its integral curves $\delta x^\gm =\der^\gm \tilde{t}\, \delta s$,
where $s$ is an arbitrary parameter. For such a displacement
$\delta x^\gm$ we have $\delta \tilde{t} =\der_\gm
\tilde{t}\,\delta x^\gm=\der_\gm \tilde{t} \der^\gm
\tilde{t}\,\delta s$. Therefore, the time-time and space-time
parts of the transformation are defined to be 
	\beq\label{all_time}
	\frac{\der x^\gm}{\der \tilde{t}}
	= \frac{\der^\gm\tilde{t}}{\der_\gn \tilde{t}\der^\gn \tilde{t}} 
	\eeq 
by eliminating $\d s$ in (\ref{normal}). We can now derive the
time-space and space-space parts. Defining $\gL^\gm{}_\tj \equiv
\der x^\gm / \der \tilde{x}^\tj\,$, consider another displacement
$\delta x^\gm$ on a $\ttt=\mathrm{const.}$ hypersurface: 
	\beq\label{tangent} 
	\delta x^\gm\der_\gm\ttt=0\,. 
	\eeq 
Writing $\delta x^\gm=\gL^\gm{}_{\ti}\,\delta \tx^\ti$ ($\delta \ttt$ is 
zero for this displacement), we find from (\ref{tangent}) for an arbitrary
$\delta \tx^\ti$ that 
	\beq\label{space-space}
	\gL^\gm{}_\tj\der_\gm\ttt=0 
	\qquad\Rightarrow\qquad
	\gL^0{}_\tj=-\frac{\gL^i{}_\tj\,\der_i \ttt}{\der_t \ttt}\,. 
	\eeq
So, we need to specify only the $\gL^i{}_\tj$ part of the
transformation. Integrating equation (\ref{all_time}) along a
$\tx^\tj= \mathrm{const.}$ curve gives 
	\beq\label{spatial_xfn}
	x^j=f^j(\tilde{x}^\ti)+\int \d \ttt \, 
	\frac{\der^j\ttt}{\der_\gn \ttt\der^\gn \ttt}\,, 
	\eeq 
with $f^j$ some arbitrary function
depending on the initial labelling of the new spatial coordinates.
Since at early times spacetime is homogeneous, useful time
slicings are those coinciding with the preferred one initially,
$\ttt=t=t_\mathrm{pref}$. Hence it is natural to consider spatial
coordinates that reduce to the preferred set on these initial time
slices. We thus take $f^j(\tilde{x}^\ti)=\delta^j{}_\ti \,
\tx^\ti$. Of course, we could perform an arbitrary spatial
transformation, but on long wavelengths it would be independent of
time, as shown in \cite{sb}. Given this result, equations
(\ref{basic zeta}) and
(\ref{constr_zeta_0})--(\ref{constr_zeta_3}) are covariant under
such a transformation. Taking a spatial derivative of
(\ref{spatial_xfn}) with respect to $\tx^\tj$, using the chain
rule and (\ref{space-space}), one arrives at 
	\beq 
	\left[\delta^j{}_l +\frac{\der_l \ttt}{\der_t \ttt} \, 
	\der_t \int \d\ttt \,\frac{\der^j\ttt}{\der_\gn \ttt\der^\gn \ttt} 
	-\der_l \int \d \ttt \,\frac{\der^j\ttt}{\der_\gn \ttt\der^\gn \ttt}
	\right]\gL^l{}_{\tk}=\delta^j{}_\tk\,, 
	\eeq 
so that 
	\beq
	\gL^i{}_\tj=\delta^i{}_\tj + \cO\left[\left(\der_i\right)^2\right]\,,
	\qquad
	\gL^0{}_\tj=-\delta^i{}_\tj\frac{\der_i\ttt}{\der_t\ttt}
	+ \cO\left[\left(\der_i\right)^2\right] \,. 
	\eeq 
The full transformation has now been specified. With these formulae one can
verify that the metric in the new coordinates reads 
	\beq
	e^{2\tilde{\ga}(\ttt,\tilde{\vc{x}})}\tilde{h}_{\ti\tj}(\tilde{\vc{x}})
	=\delta^k{}_\ti\delta^k{}_\tj\,
	e^{2\ga(t,{\vc{x}})}h_{kl}({\vc{x}})\,, 
	\qquad 
	g_{\ttt\ti}=0\,,
	\qquad 
	\tilde{N}={N/\der_t\ttt}\,. 
	\eeq 
We note that time derivatives transform as 
	\beq 
	\frac{1}{\tilde{N}}\,\der_\ttt =
	\frac{1}{{N}}\,\der_t + \cO\left[\left(\der_i\right)^2\right]\,.
	\eeq 
Under these long-wavelength time-slicing transformations the
determinant of the spatial metric, $\exp[2\ga]$, transforms as a
spacetime scalar. Hence, it is easy to see that a variable like
(\ref{general_gi}), in particular (\ref{g.i.var2}), is invariant
under these changes.

\end{document}